\documentclass[american,12pt,a4paper,numbers=noenddot,parskip=full]{cibb}
\makeatletter
\providecommand{\@ordinalM}[2]{#1}
\makeatother

\usepackage{graphicx}
\usepackage{amsmath,amsfonts,latexsym,amssymb,euscript,xr}
\usepackage{booktabs}
\usepackage[nodayofweek]{datetime}
\usepackage{hyperref}
\usepackage{fmtcount}
\usepackage[absolute]{textpos}
\usepackage{siunitx} 
\usepackage{cleveref}
\usepackage[table]{xcolor}
\usepackage{color,colortbl,tabularx}

\usepackage[american]{babel}
\usepackage[protrusion=true,expansion=true]{microtype}
\usepackage{amsmath,amsfonts,amsthm}
\usepackage{pifont}
\usepackage{subcaption}
\usepackage{xspace}

\usepackage{titlesec}
\newpagestyle{main}{
\headrule
  \sethead[][Studies of Hadron Spectroscopy at Belle and Belle~II][\usepage]
    {}{Studies of Hadron Spectroscopy at Belle and Belle~II}{\usepage}%
}
\newpagestyle{firstpage}{
  \sethead[][][\usepage]
    {}{}{\usepage}%
}
\definecolor{LightBlue}{rgb}{0.88,0.9,0.9}

\newcommand{\mbct}{\ensuremath{\tilde M_{\mathrm{bc}}}\xspace}
\newcommand{\ecmbar}{\ensuremath{\bar E_{\mathrm{cm}}}\xspace}

\title{\Large{\bf{Studies of Hadron Spectroscopy at Belle and Belle~II}}\\  \small{Presented at the 32nd International Symposium on Lepton Photon Interactions at High Energies, Madison, Wisconsin, USA, August 25-29, 2025}}

\author{\large Martin Bartl$^1$ for the Belle and Belle~II collaborations}
\address{\footnotesize $\ $\\$^1$ Max Planck Institute for Physics, Garching,
Germany\\
\phantom{$^1$ }Email: martin.bartl@mpp.mpg.de
}

\abstract{
{\bf Abstract.} The Belle and Belle~II experiments have collected an $1.6~\mathrm{ab}^{-1}$ sample of $e^+e^-$ collision data at center-of-mass energies near the $\Upsilon(nS)$ resonances. In particular, the Belle~II experiment collected a 19.2~fb$^{-1}$ sample of data at center-of-mass energies near the $\Upsilon(10753)$ resonance. We study the following processes: $e^+e^-\to \Upsilon(nS)\eta$, $e^+e^-\to \gamma X_b(\chi_{bJ}\pi^+\pi^-)$, and $e^{+}e^{-}\to\chi_{bJ}(1P) \gamma$. These results provide additional information about the nature of the $\Upsilon(10753)$ resonance and nearby structures.
In addition, we measure the $B^{0}$ and $B^+$ meson mass difference, and $\sigma\left(e^+ e^-\to J/\psi p\bar{p}\right)$  over a range of center-of-mass energies accessed via initial-state radiation.

}

\begin{document}

\thispagestyle{main}
\pagestyle{main}
\thispagestyle{firstpage}

\section{The Belle and Belle II Experiments}
\label{sec:detector_datasets}
The Belle experiment collected data between 1999 and 2010 at the asymmetric-energy $e^+e^-$ collider KEKB in Tsukuba, Japan. It was a hermetic multi-purpose detector that combined good charged-particle tracking and identification, as well as neutral detection. Combined with the clean $e^+e^-$ environment, full event reconstruction is possible. Belle's dataset, corresponding to an integrated luminosity of about $1~\mathrm{ab}^{-1}$, is the largest dataset taken by an experiment at an $e^+ e^-$ collider in the $\Upsilon$ energy range to date. Most of the data were taken with the $e^+e^-$ center-of-mass energy set to the peak of the $\Upsilon(4S)$. In addition, data were collected at the narrow resonances $\Upsilon(1S,2S,3S)$ and in an energy scan in the $\Upsilon(5S,6S)$ region.\\
The Belle~II detector comprises a major upgrade of Belle and operates at the upgraded SuperKEKB collider. Belle~II has been collecting data since 2019 and has achieved a new world-record instantaneous luminosity of $\SI{5.1e34}{cm^{-2} s^{-1}}$. The integrated luminosity reached about half of the Belle dataset. Again, the majority of the dataset has been collected at the $\Upsilon(4S)$ peak. However, 19.2~fb$^{-1}$ of data have also been taken at higher energies near the $\Upsilon(10753)$.

\section{Hadron Spectroscopy at Belle~II}
\label{sec:hadron_spectroscopy}
Hadron spectroscopy plays an important role in understanding the strong interaction, since at low energies $\mathcal{O}(\si{\GeV})$ the strong coupling $\alpha_S$ becomes too large for perturbative calculations of QCD. Especially the discovery of $X(3872)$ by Belle \cite{X3872} marked the beginning of a golden era in the discovery of exotic, i.e. non-$q\bar q$, meson states. Belle (II) has a wide-ranging physics program, including searches for new meson states and precision measurements of their properties.

\subsection{Study of $e^+ e^- \to \gamma^{\mathrm{\small ISR}} J/\psi h^+ h^-$ $ (h = \pi, K, p)$}
\label{ssec:ISR}
\begin{figure}
    \centering
    \begin{subfigure}{.49\textwidth}
    \centering
    \includegraphics[width=\textwidth]{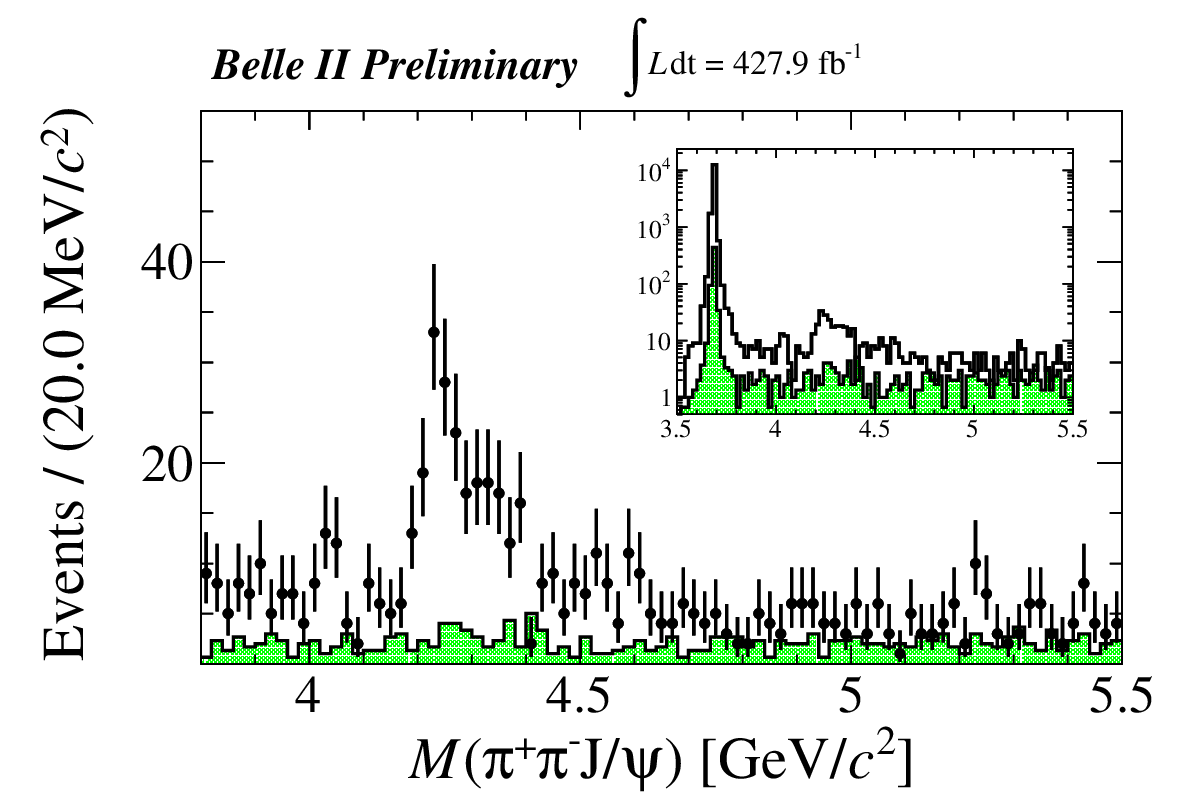}
    \caption{}
    \label{fig:pipiJpsi_mass}
    \end{subfigure}%
    ~
    \begin{subfigure}{.49\textwidth}
    \centering
    \includegraphics[width=\textwidth]{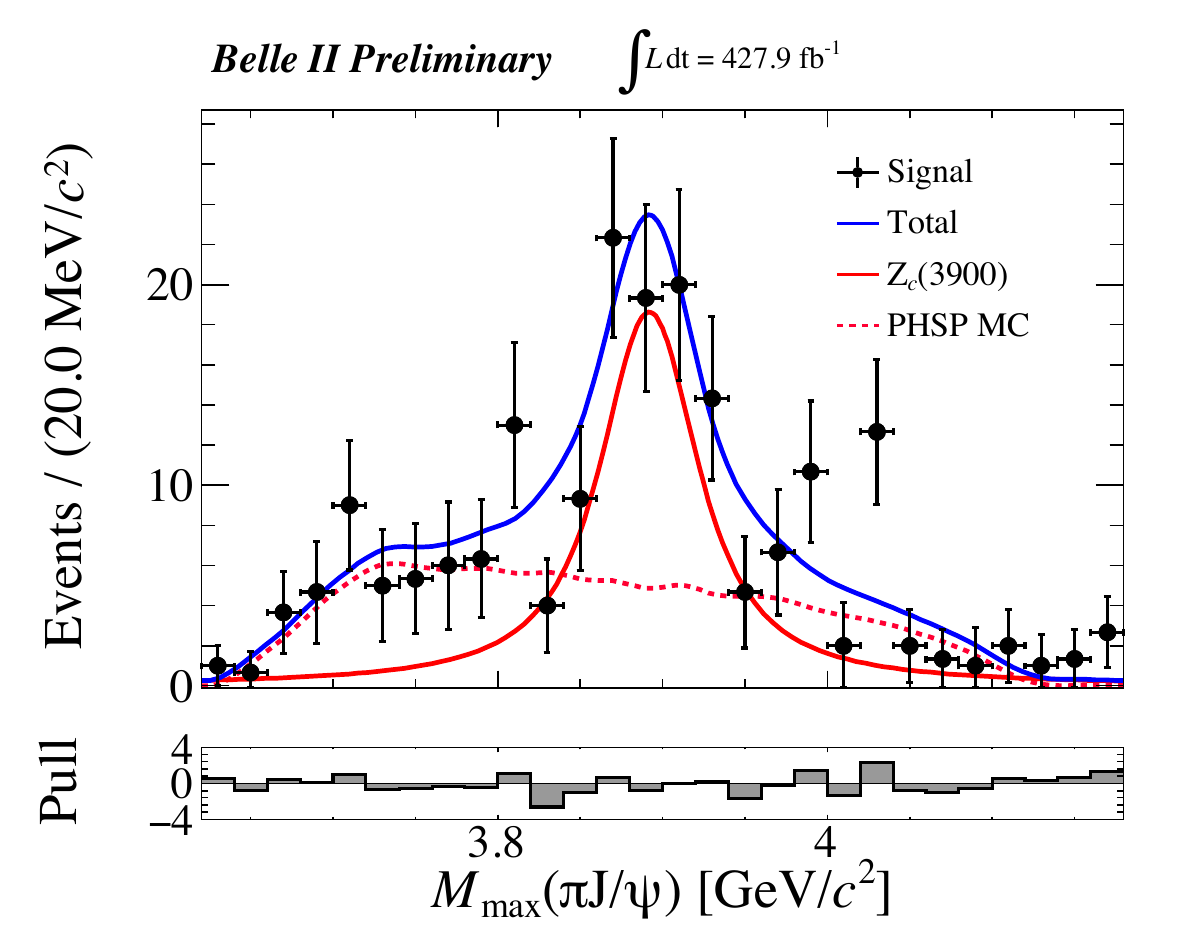}
    \caption{}
    \label{fig:fit_max_piJpsi_slim}
    \end{subfigure}%
    \caption{(a) Invariant mass distributions of the $\pi \pi J/\psi$ system. The black dots with error bars represent the selected data (signal + background), and the green shaded histograms are background estimates from the normalized $J/\psi$ sidebands. (b) Resonance-model fit to $M_\text{max}(\pi J/\psi)$ in the $\Upsilon(4260)$ $M(\pi^+\pi^-J/\psi)$ region. The black dots with error bars represent background-subtracted data, the blue curve represents the total fit, the red dashed curve represents the reweighted MC shape for pure phase-space based on BESIII result at $\sqrt{s} = \SI{4.26}{\GeV}$ \cite{BESIII_for_ISR}, and the red solid curve represents the $Z_c (3900)^\pm$ signal.
    }
    \label{fig:ISR_Plots}
\end{figure}

One open question is the nature of vector charmonium-like $Y$ states \footnote[1]{known as $\psi$ states in the PDG}, which can be beyond a simple $c\Bar{c}$ state, e.g., compact multiquarks, hadronic molecules, or hadrocharmonium. We investigate $Y$ states directly produced in initial-state radiation processes at Belle~II, e.g., on the reaction $e^+ e^- \to \gamma^{\text{\small ISR}} J/\psi h^+ h^- (h = \pi, K, p)$ with $\SI{427.9}{fb^{-1}}$ of data collected at the $\Upsilon (4S)$ resonance. We observe clear resonance structures, i.e. peaks in the invariant mass distributions, only for $h=\pi$, i.e. the $Y(4260)$ in the invariant mass distribution of $J/\psi \pi^+ \pi^-$ system in \cref{fig:pipiJpsi_mass} and the $Z_c(3900)$ in the $M_\text{max}(\pi J/\psi)$ distribution in \cref{fig:fit_max_piJpsi_slim}, where $M_\text{max}(\pi J/\psi)$ is the maximum of the invariant masses of the $J/\psi \pi^+$ and $J/\psi \pi^-$ subsystems. 
These resonance structures are in agreement with measurements from Belle \cite{JpsihhBelle} and BESIII \cite{JpsihhBESIII}.

\subsection{Observation of $e^+ e^- \to \eta \Upsilon(2S)$}
\label{ssec:etaUpsilon2S}
In addition to hadrons with $c\bar{c}$ contribution, Belle~II also searches for states with $b\bar{b}$ contribution, e.g. $\Upsilon$ states. The first observation of $\Upsilon(10753)$ by Belle \cite{UpsilonSpectroscopy} motivated Belle~II to collect additional data at four points in its energy region corresponding to 19.2~fb$^{-1}$, which is significantly larger than Belle's $\approx 6\si{fb^{-1}}$. After confirming the $\Upsilon(10753)$ in $\Upsilon(nS) \pi^+ \pi^-$ decays \cite{Belle-II:upsilon_conf}, we use the $\Upsilon(10753)$-scan data to search for other decay modes, i.e. in $e^+ e^- \to \eta \Upsilon(1S,2S)$ and $e^+ e^- \to \gamma X_b$ \cite{BelleIIpaper_eta_upsilon}.\\
We observe a signal only in $e^+ e^- \to \eta \Upsilon(2S)$ and extract the Born cross-section at four energy points. A combined fit to these cross-sections and one additional energy point from Belle \cite{Belle_additional_point} yields a significance of $< 2\sigma$ for $\Upsilon(10753) \to \eta \Upsilon(2S)$ and $>3.6\sigma$ for a new state near the $B^*\bar{B}^*$ threshold in the $\eta \Upsilon (2S)$ system, which was first described by Belle~II in \cite{Belle-II:BB_bound_state}.

\section{Measurement of the mass difference $m(B^0) - m(B^+)$}
\label{sec:DeltaM}
\begin{figure}
    \centering
    \begin{subfigure}{.49\textwidth}
    \centering
    \includegraphics[width=\textwidth]{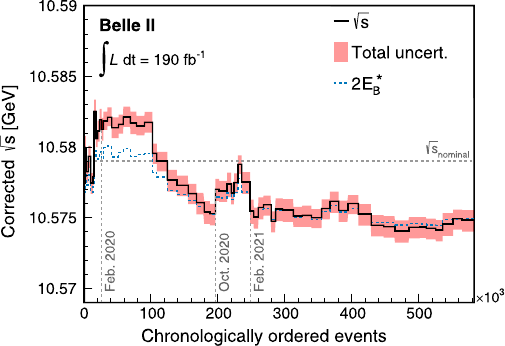}
    \caption{}
    \label{fig:ECM_over_time}
    \end{subfigure}%
    ~
    \begin{subfigure}{.49\textwidth}
    \centering
    \includegraphics[width=.8\textwidth]{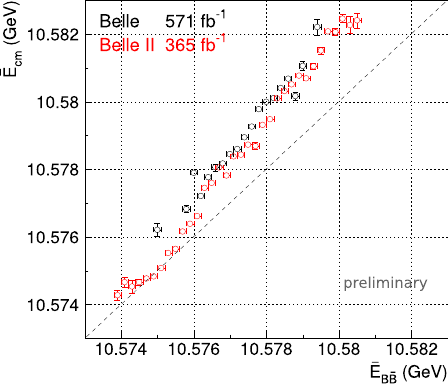}
    \caption{}
    \label{fig:ECM_vs_EBB}
    \end{subfigure}%
    \caption{(a) Average center of mass energy $\sqrt{s}$ as a function of time for data collected by Belle~II in 2020 and 2021. Taken from \cite{BelleII_CM_over_time}. (b) Average energy of the colliding beams as a function of the binned average energy of the $B\bar{B}$ pair. The black and red dots correspond to the Belle and Belle~II data, respectively. Horizontal error bars indicate $\bar{E}_{B\bar{B}}$ bins, while vertical error bars show statistical uncertainty in $\bar{E}_\text{cm}$.}
    \label{fig:Delta_M_cross}
\end{figure}

We also perform precise determinations of the fundamental properties of hadronic states. Using Belle + Belle~II data at the $\Upsilon(4S)$ ($\SI{571}{fb^{-1}} + \SI{365}{fb^{-1}}$), we measure the mass difference $\Delta m$ of $m(B^0)$ and $m(B^+)$. $\Delta m$ is an important input for quark models and constrains $m_d - m_u$.\\
To this end, we analyze the $B^0$ and $B^+$ momenta spectra, both reconstructed using a full-event-interpretation algorithm\cite{FEI}. Under the assumption that both are produced with the same energy $E_B$, we can, in principle, measure $\Delta m$ using
\begin{equation}
    \Delta m = \sqrt{E_B^2 - p_{B^0}^2} - \sqrt{E_B^2 - p_{B^+}^2}.
    \label{eq:Delta_m}
\end{equation}
The kinematics at the $\Upsilon (4S)$ are well suited for this measurement since the uncertainty of $\Delta m$ is suppressed by the small $B$ momentum as $\delta(\Delta m) \propto \frac{p_B}{E_B} \delta p_B \approx \frac{1}{16} \delta p_B$.\\
In this analysis, however, we use the distributions in
\begin{equation}
				\mbct \equiv \sqrt{\left(\frac{m_{\Upsilon(4S)}}{2}\right)^2 - p_B^2},
\end{equation}
which is uniquely related to $p_B$ of $B^0$ or $B^+$.
This choice simplifies the modeling of combinatorial background shapes. As \mbct depends on the center-of-mass energy $E_\text{CM}$ through $p_B$ in \cref{eq:Delta_m}, an accurate model of the $E_\text{CM}$ distribution is essential. Our model accounts for various effects. It models the smearing of $E_\text{CM}$ from synchrotron and initial-state radiation. Average center-of-mass energies $\bar{E}_\text{CM}$ changes over time (see \cref{fig:ECM_over_time}) and are calculated for bins of average $B\bar{B}$ energies $\bar E_{B \bar B}$ (see \cref{fig:ECM_vs_EBB}). This leads to differences between the $\bar{E}_{B\bar B}$ and beam energy spectra due to the energy dependence of the $e^+e^- \to B\bar B$ cross section, and differences between $B^0\bar B^0$ and $B^+B^-$ spectra arising from the energy dependence of $\mathcal{R}$, with
\begin{equation}
			\mathcal R \equiv \frac{\sigma(e^+e^-\to B^0\bar B^0)}{\sigma(e^+e^- \to B^+B^-)}.
\end{equation}
$\mathcal{R}$ is extracted in bins of the average center-of-mass energy \ecmbar. We parametrize the unknown energy dependence of $\mathcal{R}$ by parameterizing the energy dependence of the cross section as
\begin{align}
				\sigma(e^+e^-\to B^+ B^-) &= P_{11}(\ecmbar) p_{B^+}^3,\\
				\sigma(e^+e^-\to B^0 \bar B^0) &= P_{11}(\ecmbar) p_{B^0}^3 P_2(\ecmbar),
\end{align}
where $P_{11}(\ecmbar)$ is an 11th-order polynomial common for both cross sections and $P_2(\ecmbar)$ is a second-order polynomial that accounts for additional energy dependence of the cross-section ratio beyond the different phase-space factors $p^3_{B^0}$ and $p^3_{B^+}$.

\begin{figure}
    \centering
    \begin{subfigure}{.49\textwidth}
    \centering
    \includegraphics[width=.85\textwidth]{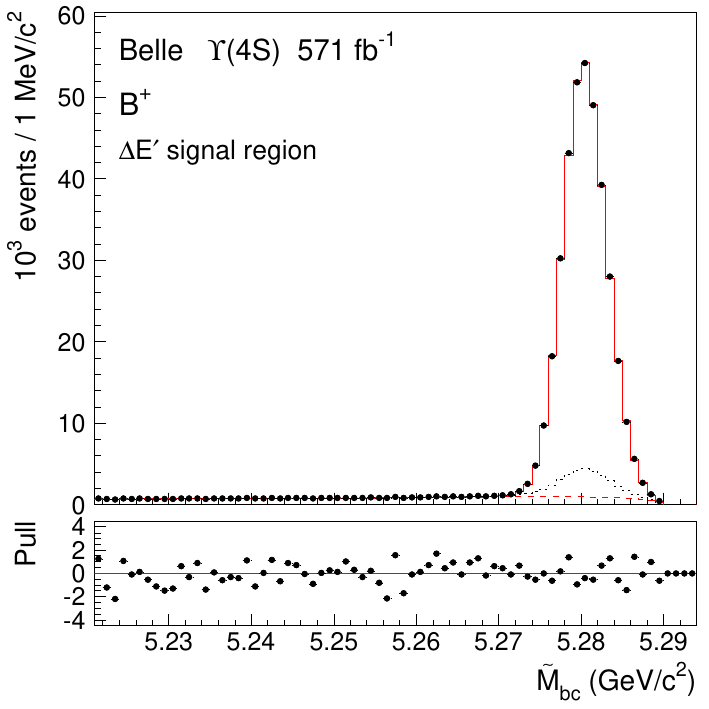}
    \caption{}
    \label{fig:mbct_fit}
    \end{subfigure}%
    ~
    \begin{subfigure}{.49\textwidth}
    \centering
    \includegraphics[width=\textwidth]{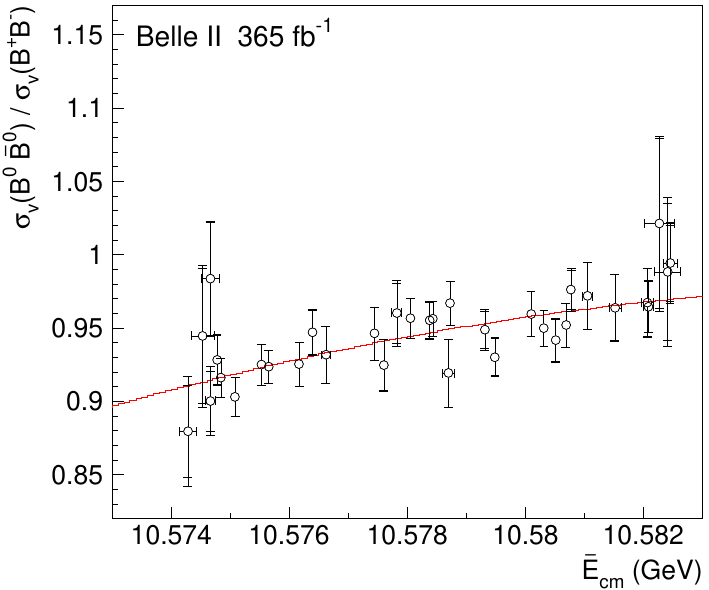}
    \caption{}
    \label{fig:R_fit}
    \end{subfigure}%
    \caption{(a) \mbct distribution for $B^+$ from Belle shown as black dots. The red solid line represents the fit, the red dashed line the background contribution, and the black dashed line the broken signal, i.e. signal events where one of the final state particles originates from background. (b) Energy dependence of the ratio $\mathcal{R}$ of visible cross sections. The dots with error bars show the direct measurement performed using Belle II data. The vertical inner and outer error bars show statistical and total uncorrelated uncertainties, respectively. The horizontal error bars show statistical uncertainty. The red curve represents the fit result.}
    \label{fig:combined_fit}
\end{figure}

Finally, we measure $\Delta m$ by performing a combined fit to 8 \mbct distributions, i.e. Belle and Belle II data, each for $B^0$ and $B^+$ in a signal and sideband region (see exemplary \cref{fig:mbct_fit}), as well as $\mathcal{R}$ distributions for Belle and Belle II (see exemplary \cref{fig:R_fit}). As an additional constraint, we also fit to the energy dependence of $\sigma(e^+e^-\to b\bar b)$ measured by BaBar \cite{Babar_cross_section} and $\sigma(e^+e^-\to b\bar b\to D^0/\bar D^0 X)$ measured by us. We extensively test the systematic effects, including the binning in \mbct, and the $p_B$ resolution. We also test the parameterizations used in the combined fit, e.g. the choice of polynomial order of $P_{11}(\ecmbar)$ and $P_2(\ecmbar)$, as well as the background modeling. One key finding is that the precision deteriorates significantly if either Belle or Belle~II data on $\mathcal{R}(\ecmbar)$ is excluded from the fit.

We measure
\begin{equation}
    \Delta m = (0.495 \pm 0.024 \pm 0.005)\,\text{MeV}/c^2,
\end{equation}
which is the most precise measurement of $\Delta m$ to date. It differs significantly from the BaBar \cite{BaBar_deltam} result, $(0.33 \pm 0.05 \pm 0.03)\,\text{MeV}/c^2$, which dominates the current PDG average. 
To study the origin of this difference, we perform a combined fit using the phase-space hypothesis, $\mathcal{R} = (p_{B^0} /p_{B^+})^3$ as assumed by BaBar. This results in $\Delta m = (0.386\pm0.006)\,\text{MeV}/c^2$, which is consistent with the BaBar value. However, this fit describes the data significantly worse. Hence, the pure phase-space hypothesis is disfavored at approximately $10 \sigma$ according to Wilks' theorem, which confirms predictions from \cite{bias_for_BaBar_theory} that this assumption could lead to a bias in $\Delta m$.
In addition, we extract the energy dependence of $\mathcal{R}$ in the combined fit for phenomenological studies. We perform a phenomenological analysis of the extracted $\mathcal{R}(\ecmbar)$ to determine the parameters of the isovector $B\bar B$ potential, which modifies $\mathcal{R}(\ecmbar)$. We find a $2\sigma$ indication that the isovector $B\bar B$ potential is attractive.

\section{Summary}
\label{sec:conclusion}
In summary, Belle and Belle~II have rich physics programs in the field of hadron spectroscopy, utilizing their unique datasets. We measured $e^+ e^- \to \gamma^{\mathrm{\small ISR}} J/\psi h^+ h^-$ $ (h = \pi, K, p)$ study $Y$ and $Z$ states, where we find significant resonance structure only in $e^+e^-\to \gamma^{\mathrm{\small ISR}} J/\psi \pi^+ \pi^-$, i.e. a $Y(4260)$ signal in the $J/\psi \pi^+ \pi^-$ system and a $Z_c(3900)$ signal in the $J/\psi \pi$ subsystem.
In addition, we observed $e^+ e^- \to \eta \Upsilon(2S)$ at $E_{CM}$ around the $\Upsilon(10753)$ mass, where a fit to the measured cross-section shows no significance of $\Upsilon(10753)$ signal and evidence for a possible state near the $B^*\Bar{B}^*$ threshold at a $> 3.6\sigma$ level.
Finally, we measure the mass difference $m(B^0) - m(B^+)$ with world-leading precision, providing important input for quark models.
In the future, Belle~II will collect even larger data sets, enabling us to study hadron spectroscopy at Belle and Belle II with unprecedented precision.

\printbibliography

\end{document}